\documentclass[%
 aip,
 jmp,
 amsmath,amssymb,
preprint,%
]{revtex4-1}

\usepackage{graphicx}% Include figure files
\usepackage{dcolumn}% Align table columns on decimal point
\usepackage{bm}% bold math
\usepackage{gensymb}
\usepackage[colorinlistoftodos]{todonotes}

\usepackage{epstopdf}
\usepackage[english]{babel}
\usepackage[utf8]{inputenc}

\begin{document}

\title{Flow sensing by buckling monitoring of electrothermally actuated  double-clamped micro beams}

\author{Y. Kessler, S. Krylov, A. Liberzon}
\affiliation{School of Mechanical Engineering, Tel Aviv University, Tel Aviv 69978, Israel}

\date{\today}% It is always \today, today,

\begin{abstract}
We report on a flow sensing approach based on deflection monitoring of micro beams buckled by the compressive thermal stress due to electrothermal Joule's heating.
The  air stream convectively cooling the device affects both the critical buckling values of the electric current and  the postbuckling deflections of the structure.  After calibration, the flow velocity was obtained from the deflections measurements.
 The quasi-static responses of 2000 $\mu$m  long, 10 $\mu$m wide and 30 $\mu$m high single crystal silicon beam transduced using image processing  were consistent with the prediction of the reduced order model, which couples thermoelectric, thermofluidic and structural domains. The deflection sensitivity of 1.5 $\mu$m/(m/s) and the critical current  sensitivity of  0.4 mA/(m/s) were registered in the experiments. Our model and experimental results collectively demonstrate feasibility of the sensing approach and  further suggest that simple, robust and potentially downscalable beam-type devices may have use in flow velocity and wall shear stress sensors.
  % for miniature unmanned air vehicles, environmental, biometric and personal health care applications.
 \end{abstract}

%Use showkeys class option if keyword
\maketitle

\section{Introduction}
\label{sec:intro}

Minimally invasive flow velocity and wall shear stress (WSS) measurements are necessary in various areas of science and engineering  \cite{tavoularis2005measurement} starting from flow control in pipelines,  aerodynamics of lifting bodies and wind turbines, lab-on-a-chip and micromechanical-valves development \cite{Tavakol2016,Ducloux2007}, biomedical and personal health care applications and up  to fine scientific instruments for fundamental fluid dynamics research~\cite{borisenkov2015}.
Microelectromechanical systems (MEMS) based flow sensors are advantageous due to their small size, low cost and power consumption, high sensitivity\cite{ho1998micro} and resolution~\cite{lofdahl1999mems, wang2009mems}, and integrability with electronic circuitry.
The most common free shear flow sensors are based on the temperature dependent electrical resistance of overheated thin films or wires~\cite{stainback1993review, borisenkov2015},  suspended or attached to a substrate.
Despite a significant research effort devoted to miniaturization of flow measurement devices~\cite{gad1989advances, wang2009mems},
many challenges still have to be addressed. Flow micro sensors in wall-bounded flows, following either mechanical (e.g. floating element \cite{schmidt1988design}) or thermal (e.g. heat transfer \cite{jiang1994theoretical,kalvesten1996pressure, jiang1996surface}) measurement principles\cite{gad1989advances,wang2009mems}, are typically distinguished by intricate design and complex fabrication process.
\begin{figure}[t]
\includegraphics[width=.9\columnwidth]{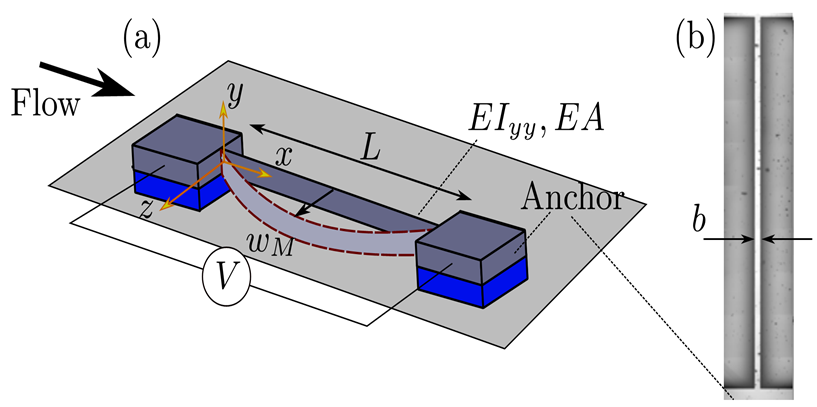}
\caption{(a) Schematics of the device: a double clamped beam is attached to the substrate by two anchors and is aligned parallel to the flow direction. (b) optical microscope image (top view) of a fabricated beam, $b = 9.7\pm 0.3\;\mu$m \label{fig:beam}}
\end{figure}

In this work we introduce a flow measurement approach based on the static deflection of the double-clamped micro beam buckled by a compressive stress, which is originated in the electrothermal actuation by
Joule's heating. The device shown in Fig.~\ref{fig:beam} combines a mechanically moving MEMS structure and an anemometer-style architecture of a flow sensor and incorporates a micro beam attached to the substrate by two anchors.  
{The beams are designed to deflect parallel to the substrate in the in-plane ($z$)  direction,} while the axial $x$ displacement of the beam's ends is fully contained by the fixed anchors. The nominally $L=2000\;\mu$m long,
$b=10\;\mu$m wide and $d=30\;\mu$m high single crystal
silicon beams aligned in the $\langle 110 \rangle$ crystallographic direction are fabricated from silicon on insulator
(SOI) substrates with (100) upper surface and $30\;\mu$m thick device layer using deep reactive ion etching (DRIE) based process~\cite{Medina2014323}. Devices are electrically isolated from
the handle by an underlying 3 $\mu$m thick silicon dioxide layer. A voltage $V$ between the anchors of the beam is the source of the electric current and consequently of the resistive (Joule's) heating of the beam. Above certain value of the current the resultant compressive thermal stress exceeds the critical, buckling, value~\cite{kaajakari2009small} and the beam bends. Conductive and convective heat transfer accompanying the air flow results in the cooling of the structure and in the decrease of the axial stress and of the deflection. Possible scenarios for the measurement of the free flow velocity or WSS  (if installed on the wall) include {registering} of the critical buckling value of the electric current/voltage or monitoring of the post buckling deflection of the beam at a constant prescribed current/voltage above the critical value. Note that while various flow directions can be applied in this work, {in order to eliminate the effect of the direct actuation of a beam by a flow,} we focus on the flow parallel to the beam axis, Fig.~\ref{fig:beam}.

{Note} that micro sensors based on a double-clamped beam architecture are distinguished by simplicity, robustness, and can be readily downsized to the nano scale.
They are 
%considered as a 
core element in many applications such as radio-frequency switches \cite{robert2003integrated}, resonant sensors \cite{sibgatullinexcitation}, sliding plate micro-valves \cite{maluf2004introduction}, gas sensors~\cite{Joe2012} as well as pressure gauges, gyroscopes, and accelerometers~\cite{tanaka2007industrial}.
Here we suggest to extend the use of beam-like devices to flow sensors. In addition, results of this work may shed light onto the interaction between such devices and an ambient flow.  

\section{\label{sec:model}Model}

We assume that the deflections $w(x)$ of the slender Euler-Bernoulli beam, while comparable with its
thickness, are small with respect to the
length of the beam and satisfy to the equilibrium equation~\cite{villaggio2005mathematical}
\begin{equation}
%E I w''''  -  \left[ N_0 + \frac{E A}{2L} \int_0^L (w')^2dx  \right] w'' = 0 \label{EQ:beamDef}
E I_{yy} w''''  +  \left[  E A \alpha \overline{\theta} - \frac{E A}{2L} \int_0^L (w')^2dx  \right] w'' = 0
%E I_{yy} w''''  -  \left[ \sigma_0 A - E A \alpha \overline{\theta} + \frac{E A}{2L} \int_0^L (w')^2dx  \right] w'' = 0
\label{EQ:beamDef}
\end{equation}
Here $(\;)'=d/dx$, $E$ and $\alpha$ are   Young's modulus (in the $\langle 110 \rangle$ direction) and coefficient of the thermal expansion of Si, respectively; $A=bd$ and $I_{yy}=db^3/12$ are, respectively, the  area and the second moment of area of the uniform cross section and $\overline{\theta} = \int_0^1 T(x)dx/L- T_\infty$ is the averaged temperature excess  above the ambient.  
While single crystal Si beams may manifest small residual stress~\cite{Medina2014323}, we assume the beam  to be initially stress-free. In accordance with Eq.~(\ref{EQ:beamDef}) the axial force consists of the compressive thermal force
as well as the nonlinear tensile stretching force, arising due to the axial constrain. 
%The integral term is the non-linear stretching related to the doubly clamped configuration\cite{emam2004nonlinear}. The compressive force, $N_0$, is given as: $ N_0= \sigma_0 A + E A \alpha \overline{\theta}$, where $A$ is the beam cross-section area, $\alpha$ is the thermal expansion coefficient and $\overline{\theta} = \overline{T - T_\infty}$ is the averaged temperature difference between the beam and the surrounding fluid.
%We will assume that the pre-stress in the beam, $\sigma_0 $, is zero. The compressive axial force term ($N_0$) dependence on $\overline{\theta}$ is due to coupling between the mechanics and the heat transfer.
%
By using the solution of Eq.~(\ref{EQ:beamDef})~\cite{nayfeh2008exact} the mid-point deflection of the beam in the postbuckling state can be obtained
\begin{equation}
% w_m =  \frac{4r}{\sqrt{3}} \,  \sqrt{\frac{\alpha \overline{\theta}}{3\epsilon_E} - 1}, \quad (\alpha \overline{\theta} > \sigma_E)
w_m =  \frac{4r}{\sqrt{3}} \,  \sqrt{\frac{\alpha \overline{\theta}}{3\epsilon_E} - 1}, \quad (\alpha \overline{\theta} > 3\epsilon_E)
\label{eq:Buckling-Solution}
\end{equation}
%
% \yoav{We didn't say anything about $\sigma_E$, should be $3\epsilon_E$?}
where $r=\sqrt{I_{yy}/A}$  and $\epsilon_E = 4\pi^2  r^2 / L^2 = N_E/EA$ is the axial strain corresponding to the critical Euler buckling force $N_E$.

The temperature excess $\theta(x)=T(x)-T_{\infty}$ is calculated
%$\overline{\theta}$ can be estimated
using the one-dimensional heat transfer equation
\begin{equation}
\kappa  \, \theta'' - \frac{h  P}{A} \, \theta + \rho_e J^2  = 0
\label{eq:heat_transfer}
\end{equation}
where $h$ is the convection heat transfer coefficient, $\kappa$ is the thermal conductivity, $P=2(b+d)$ is the beam cross-section perimeter responsible for natural convection (or $b+2d$ in the case of forced convection), and the last term is the Joule's heating source ($J=I/A$ and $\rho _{e}$ are the current density and electrical resistivity respectively). The radiation and the heat conduction to the substrate through the air gap are neglected~\cite{glassbrenner1964thermal,pelesko2002modeling}. 
%
%for constant material properties \cite{glassbrenner1964thermal},
%and it is assumed  that the heat flux due to convection is from the envelope of the beam.
%one-dimensional axial
%\sout{conduction and convection are solely from the upper and side surfaces  of the beam.Note that despite that  there is a small 2 $\mu$m gap between the beam and the substrate, it is too small for the flow to develop~\cite{pelesko2002modeling} yet too large for conductive heat losses to the substrate.}
 We also assume that at the clamped ends (the anchors) of the beam
 %and the substrate
 %are at the ambient temperature and therefore
 $\theta(0) = \theta(L) = 0$. All the material properties $E,\;\alpha,\;\kappa,\;\rho_e$ appearing in Eqs.~(\ref{EQ:beamDef}) and~(\ref{eq:heat_transfer}) are assumed to be temperature-independent and are shown in table \ref{tab:properties}. This assumption is justified since {the air flow cooling related} temperature changes, which {are} of primary interest here, are small. According to our model, for the 2 mm beam, $\overline{\theta}_{I=4.42 \mathrm{mA}} = 106.82^{\circ}$C  for 0 m/s air speed and  $\overline{\theta}_{I=4.42\mathrm{mA}}=100.83^{\circ}$C for 2 m/s, where $I=4.42$ mA is 1.1 times the critical current for the highest air velocity {used in} the experiments. In addition, we assume that the Joule's heating is dominant and the additional heating due to mechanical strain is negligible. Finally, we neglect the thermoresistive and piezoresitive effects.

Under these simplifying assumptions Eq.~(\ref{eq:heat_transfer}) is uncoupled from Eq.~(\ref{EQ:beamDef}) and can be readily solved in terms of $\overline{\theta}$~\cite{mills1999basic,holman1986heat}
\begin{equation}
%\theta(x)=\frac{\rho_e J^2}{km^2}\left[1-\cosh(mx)+\sinh(mx)\frac{\cosh(ml)-1}{\sinh(mL)}\right]
\overline{\theta}=\frac{\rho_eJ^2L^2}{4\kappa m^2}\left(\! 1-\!\frac{\tanh(m)}{m}\right), \;\; m=\frac{L}{2}\sqrt{\frac{Ph}{\kappa A}}
\label{eq:thermal_colution}
\end{equation}
%
% which is parameterized by $h$. 
The dependence of $h$ on the flow velocity $u$ results in the coupling between $u$ and the deflection of the beam through the thermal stress.
%The solution of this equation is given in textbooks\cite{mills1999basic,holman1986heat} in the form of average $\overline{\theta}$ (with boundary conditions of $\theta_{x=0} = \theta'_{x=L/2} = 0$.)  The coupling between flow velocity (the measured quantity) and temperature $\overline{\theta}$ is within the convection coefficient, $h$.

We focus on two different scenarios - forced convection due to a flow along the beam 
%length 
and natural convection at zero air velocity. 
Mixed convection scenario was estimated to be irrelevant~\cite{mills1999basic}. Due to the miniature size of the device we use the correlations for the laminar flow regimes for small Rayleigh and Reynolds numbers. For natural convection we use\cite{holman1986heat}  $h=\Sigma C_{i} \left({\overline{\theta}\, P/ (\,A_{s}\,)}\right)^{1/4}$ ($A_{s}$, which is the area of convection, and $C_i$ depend on wall orientation, for the horizontal facing upward $C_{i}$ is $1.32$ and for the vertical walls $1.42$, for the horizontal face pointing downward we use $0.59$)    while for the forced convection~\cite{whitaker1972forced} $h=Nu \; \kappa_f/L$, where  Nusselt number based on beam length is $ Nu = 0.664(Pr)^{1/3} Re^{1/2}$, and the Reynolds number $Re = u L/\nu$ (where $\kappa_f$ is the air thermal conductivity and $\nu$ is the air kinematic viscosity~\cite{holman1986heat}).

%Following Nayfeh and Emam \cite{nayfeh2008exact} we solve Eq.~\ref{EQ:beamDef} for the mid-point deflection, $w_m =  4/\sqrt{3} \, r \sqrt{N_0/N_E - %1}$ ,($r=\sqrt{I/A}$ is the radius of gyration, $N_E = 4\pi^4 E I / L^2 $ is the Euler force). High temperature difference $\overline{\theta}$ will %cause $N_0$ exceeding $N_E$ followed by buckling of the beam.

\section{Results}
% \subsection{Model Results}
%
%Combining the beam and the heat transfer equations in a coupled multi-physics model 
First we investigated the feasibility of the sensing principle using the model. 
\begin{table}[t]
\centering\caption{Properties of  single crystal silicon~\cite{glassbrenner1964thermal, okada1984precise} and of standard air (at room temperature~\cite{holman1986heat})  used in calculations \label{table1}}

\begin{tabular}{| l | c |}
\hline
Parameter & Value \\ \hline
 Young modulus, $E$ & 169 GPa \\
Thermal expansion, $\alpha$ & $3.28 \times 10^{-6}\; 1/()^{o}C$ \\ 
Thermal conductivity, $\kappa$ & $130.67$ W/(mK) \\ 
Resistance, $R$ &  1.01  K $\Omega$ \\ 
Prandtl number (air), $Pr$ &  0.7 \\
Thermal conductivity (air), $\kappa_f$ & 0.0257 W/(mK) \\
Kinematic viscosity (air), $\nu$ & $1.5 \times 10^{-5}$ m$^2$/s \\ \hline
\end{tabular}%
\label{tab:properties}
\end{table}
%
%textcolor{red}{We use properties of standard air at room temperature \cite{holman1986heat} and that of single crystal silicon \cite{glassbrenner1964thermal, okada1984precise} for the beam.}
%
%A constant voltage, $V$, was applied between the anchors and the current, $I$, was \textcolor{blue}{measured by ammeter}. 
Critical temperature (which is the temperature in which the beam buckles) was chosen to be the "working point" and used in order to set the material properties. Since the resistivity $\rho_e$ is temperature and strain dependent\cite{park2009review}, it was estimated by directly measuring the resistance $R$ for each beam at the working point (see Table~\ref{table1}) and then using the expression $\rho_e = R\, A/L$. Because the critical temperature and the material properties are interrelated (Eq.~(\ref{eq:Buckling-Solution}) and~(\ref{eq:thermal_colution})) eq.~(\ref{eq:thermal_colution}) was solved iteratively  until convergence {providing the working point values of the material properties.} Thermal conductivity-temperature and thermal expansion-temperature empirical correlations obtained from literature \cite{glassbrenner1964thermal, okada1984precise} {were used}. 
%\marginpar{\tiny{you cannot directly use (4), the equation is nonlinear! I do not understand what you did here. Did you solve the nonlinear(3) or used (4) and assumed that $\rho$ is constant?-\textcolor{blue}{$\kappa and \alpha$ explained now. earlier we show that we found R empirically using ohm's law (a linear fit of the $V=f(I)$ correlation). This is how we set it as constant. }}}
Finally, the midpoint deflection $w_m$ of the beam was obtained using Eq.~(\ref{eq:Buckling-Solution}).
% Although the technique is less accurate than the four-probe method~\cite{smits1958measurement} for better precision).   % Note that since $h$ in the case of natural convection depends on $\overline{\theta}$, the problem is nonlinear and was solved using an iterative procedure.

\begin{figure}
% \subfloat[]{\includegraphics[width=.8\columnwidth]{model_1mm.eps}}
% \subfloat[]{\includegraphics[width=.8\columnwidth]{model_2mm.eps}}\\
%\subfloat[]{\includegraphics[width=.8\columnwidth]{exp_1mm.eps}}
% \subfloat[]{\includegraphics[width=.8\columnwidth]{exp_2mm.eps}}
\includegraphics[width=.9\columnwidth]{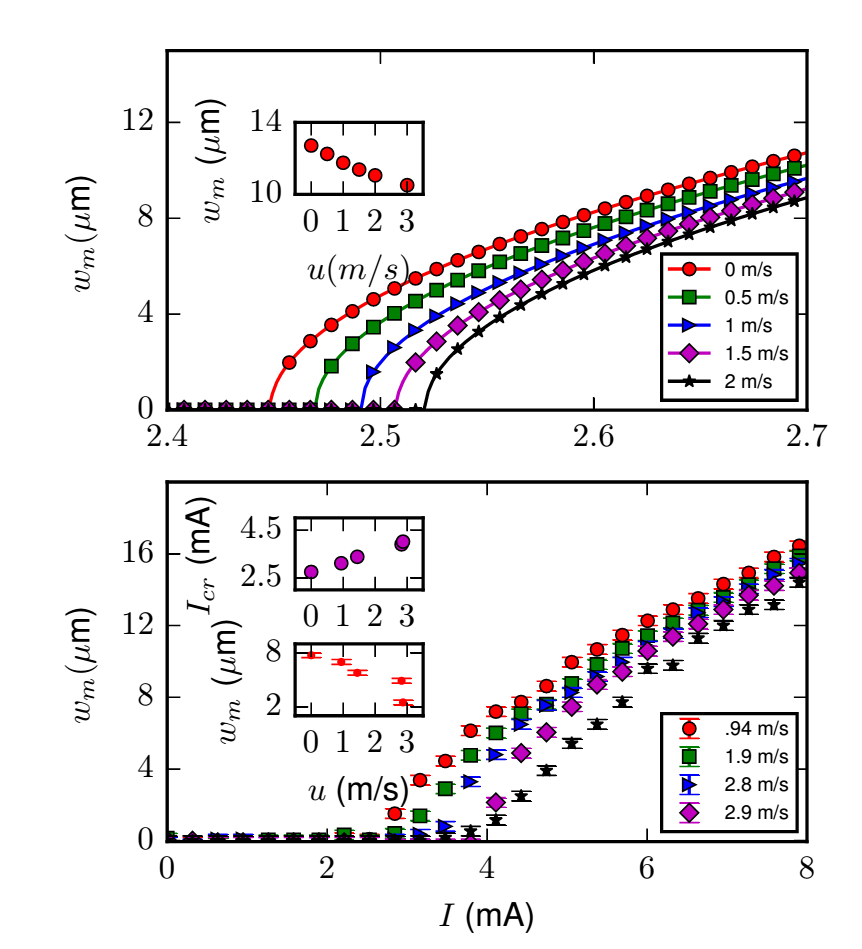}
\caption{Model results (top) and  experimental results (bottom) for increasing air velocity from left to right. Insets show post-buckling deflection at constant current of 1.1 the critical current (4.42 mA and 2.8 mA for the experiment and model respectively) as a function of flow velocity and of critical current for buckling as function of flow velocity.% \yoav{add legend. error bars should be constant. switch between inserts (post buckling higher and blue) and start from -0.01 x axis}
}
\label{fig:resultsout}
\end{figure}

Results of calculations, performed for the nominal geometry of the beam, are shown in Fig.~\ref{fig:resultsout} (top). At a certain critical current $I_{\mathrm{cr}}$ the beam buckles and the deflection  sharply changes from zero (pre-buckling) to a post-buckling value. Due to the cooling by the forced convection higher current is required to reach the critical value of the temperature and of the stress. Consequently,  $I_{cr}$ increases with the flow velocity. Further increase of the current above $I_{cr}$ is accompanied by the increase of the deflection.
As expected the deflection of the buckled beam at a constant voltage/current decreases as the flow velocity increases (see inserts in Fig.~(\ref{fig:resultsout})). 
%also demonstrates the dependence of the bucking point ($I_{\mathrm{cr}}$) on the external flow velocity, that varies from zero (natural convection) to the forced convection cases of increasing flow velocity parallel to the beam.
%
% \captionsetup[subfigure]{position=top, labelfont=bf,textfont=normalfont,singlelinecheck=off,justification=raggedright}
% \captionsetup[subfigure]{position=top,textfont=normalfont,singlelinecheck=off,justification=raggedright}
%

% \section{Experiment}\label{sec:Exp}
% \subsection{MEMS double-clamped beam design}
%
%We test the feasibility of the proposed design in its microscopic form using a double-clamped beams fabricated from silicon on insulator (SOI) wafer %by deep reactive ion etching (DRIE).
% \subsection{Test rig}
%

The beams of various configurations were mounted on a wafer prober (Karl Suss PSM-6 with FS-70 Mitutoyo microscope) and tested in ambient air conditions. Note that while large number of beams of differing geometries (cross-section, length, aspect ratio) were operated, here, for the sake of brevity, we present the results of only one.
%We have tested several beam designs, studying the effect of those parameters on the measurement sensitivity of the device. For the sake of brevity we demonstrate here one set of the results, as 
All beams behave similarly in terms of their response to different flow velocities in our test rig.

For the buckling  experiments, the AC current is supplied by a signal generator (TGA1241,  Aim-TTi, England) that generated a triangular wave with time period of 60 seconds and peak-to-peak voltage of 10 Volts. For post-buckling mode of operation, the DC voltage of 4.47 Volts was supplied (EX752M , Aim-TTi, England) and the current was registered by a multi-meter (M-3890D, METEX, Korea)  connected in series with the beam. Beam deflection was recorded using a CCD camera (UI-2250SE-M-GL, iDS GmbH) equipped with a $\times$50 lens. The video was post-processed with custom-made image processing code using common methods.
Dry air was supplied through a pressure regulator/gauge and filters using a long straight needle (inner diameter 1 mm, length 150 mm). The pressure-controlled flow produced by a custom built miniature blower was calibrated using particle imaging velocimetry (PIV).
%

% \textbf{UP TO HERE}
%We have tested several beam designs (cross-section, size, aspect ratio factor, etc.), studying the effect of these parameters on the measurement sensitivity of the device. For the sake of brevity we demonstrate here one set of the results, as all beams behave similarly in terms of its response to different flow velocities in our test rig. We compare two beams of different lengths with the same width and height and demonstrate the effect of this aspect ratio on the measurement resolution, as shown in Fig.~\ref{fig:resultsout} (bottom panel). 
%
In Fig.~\ref{fig:resultsout} (bottom) we show the results of the buckling experiment namely the measured deflection ($w_m$) versus the supplied current ($I$), for different flow velocities.  
%In the insets of each figure we present the results of the post-buckling experiments and of the critical current for buckling as function of flow velocity. 
Error bars emphasize the measurement uncertainty, which is mainly due to the resolution of our imaging system. We observe stronger critical buckling currents with increasing flow velocity. It is expected as the flow that cools the device requires stronger Joule's heating for the beam to buckle. In the lower inset we see the same effect from a different perspective - when the constant voltage of 4.47 Volts is applied, the deflection of the post-buckled beam decreases with the increasing flow velocity, as expected from the model. In the upper insert we demonstrate the increase in critical current as flow velocity grows. 

The experimental results show clearly the behavior predicted by the model and the expected response of the deflection to the flow velocity (Fig.~(\ref{fig:resultsout})). The results corresponding to different beams are very similar qualitatively and differ quantitatively due to
fabrication-related uncertainty in device's parameters as well as due to  a long list of parameters and effects that were neglected or simplified in the model (Sec. \ref{sec:model}). %, such as, for instance, thermo-resistivity of piezo-resistive single crystal silicon\cite{park2009review}.

\begin{figure}
\includegraphics[width=.9\columnwidth]{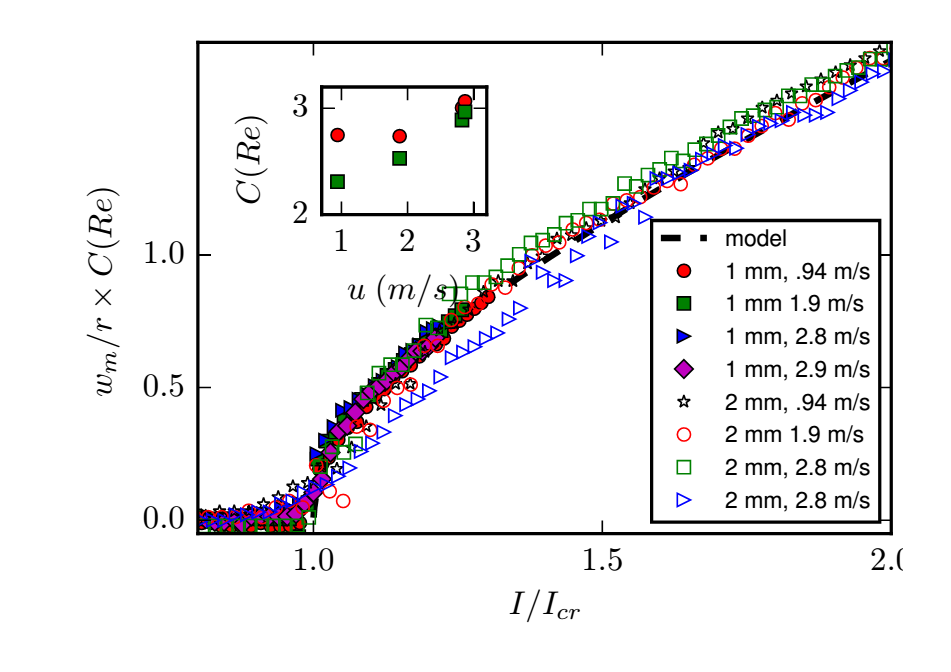}
\caption{Dimensionless presentation of the experimental results for the two beam lengths and different flow velocities. The model curve corresponds to the $w_m/r = C(Re) \sqrt{(I/I_{cr})^2 - 1}$. Inset shows $C(Re)$ constant for different velocities and the two beam lengths (circles $L=1$ mm, squares $L = 2$ mm). \label{fig:dimensional}}
\end{figure}

In order to understand better the dependence of the system response on various parameters we analyze the model based on dimensional grounds. This analysis allows predicting the deflection and the critical current for double-clamped beams of different lengths and at different flow velocities. Using the result of this analysis in conjunction with the model we can also deduce the constants that effectively represent the effects neglected in the simplified model such as temperature dependence of material's properties, convection and flow regime issues, uncertainties in geometry, residual stress, piezo-resistive effects \cite{park2009review} and others. In Fig.~\ref{fig:dimensional} the experimental results of different beams are compiled into a single, self-similar, curve presented in the dimensionless form. We demonstrate that a single constant which depends on Reynolds number is sufficient to represent all our results. From Fig.~\ref{fig:dimensional} it is clear that the model represents well the dynamics of the method (device) for the current which is up to at least two times the critical (buckling) value, $I \le 2 I_{\text{cr}}$. %The scatter of the results at higher values point out the discrepancy with the simplified flow-heat transfer correlations. 
Furthermore, the model that collapses the data points out that the heat transfer correlations has to be taken with care, especially those based on the assumptions of fully developed laminar flow parallel to the device. In any case, this result  does not invalidate our analysis, but strengthens the comprehensive analysis of the complex dynamics of the flow along the buckling beams.
%

% \subsection{Sensitivity}\label{subsec:sensitivity}
%

Sensitivity (or scale factor) of a sensor can be predicted using the model and quantified specifically for a given sensor empirically. From model and experimental results, we understand that sensitivity improves mainly for slower flow velocities, smaller deflections (or lower temperatures), and lower 
%{bending stiffness}, 
buckling force, $\propto E I /L^2$. There is, however, a natural trade-off between sensitivity and robustness of the sensor.  For our devices we obtain sensitivity better than majority of the reported MEMS sensors\cite{wang2009mems}:  0.44 and  0.435 V/(m/s) (or 0.41 and 0.43 mA/(m/s))  for shorter and longer beams, respectively, estimated at critical point of buckling. In addition, for the post buckling experiments (measured at 12.25 and 4.42 mA), we get scale factors of 0.41 and 1.5 $\mu$m/(m/s) for shorter and longer beams, respectively.

\section{Conclusions} \label{sec:conclusions}

In this work, the novel measurement concept for the MEMS-based double-clamped beam flow sensor is proposed, designed, analyzed and tested. We introduce a sensing method, based on coupling between different physical domains including thermal, flow, mechanical and structural.  The innovative aspects of the sensor are a) rather complex but well understood non-linear, multi-physical, method that couples Joule's heating, mechanical motion of the beam, flow, convective and conductive heat transfer, b) relatively low heat transfer to the substrate, c) low cost and straightforward fabrication, d) can be downscaled to nanometer size, e) very good sensitivity, and f) versatile modes of operation as the device can be built and installed on the wall or in the free flow setup (anemometer type), it can be used in buckling and post buckling methods. The model of the fluid-structure interaction due to convective heat transfer, Joule heating and double-clamped beam buckling is studied in the simplified form, yet numerical simulation based on this model is in a very good agreement with the experimental results of a MEMS device in a custom-designed test rig. Furthermore, similarity analysis based on dimensional grounds emphasizes better the response of the real device to the flow parallel to the beam length, explains the discrepancies between the model and the experiments and suggests the improved design for sensitivity and robustness of the measurement method.

% There are several possible drawbacks in the proposed design: since the measurement method is indirect, it provides the velocity measurements based on the correlations that could be achieved only through calibration. There is a need for additional study to understand how the sensor will respond to the changes in the flow pattern such as reverse flows and separations, laminar to turbulent transitions, change in temperature and others.  Future work will be devoted to further develop this measurement concept. We will also study the option to create a parametric resonant motion of the buckled beam and develop a dynamic sensor, based on the proposed model combined with additional inertial and dumping expressions.

\section{Acknowledgments}

The devices were fabricated  at the Tel Aviv University Micro and Nano central characterization and fabrication facility (MNCF). The authors would like to thank Dr. Sasha Gurevich, Naftaly Karakover and Erez Benjamin for their help with fabrication;  Stella Lulinski and Eli Kronish for their help with experiments and experimental-setup, respectively.

\bibliographystyle{apsrev4-1}%{apl}%{aipnum}
\bibliography{YKref}

\begin{thebibliography}{29}%
\makeatletter
\providecommand \@ifxundefined [1]{%
 \@ifx{#1\undefined}
}%
\providecommand \@ifnum [1]{%
 \ifnum #1\expandafter \@firstoftwo
 \else \expandafter \@secondoftwo
 \fi
}%
\providecommand \@ifx [1]{%
 \ifx #1\expandafter \@firstoftwo
 \else \expandafter \@secondoftwo
 \fi
}%
\providecommand \natexlab [1]{#1}%
\providecommand \enquote  [1]{``#1''}%
\providecommand \bibnamefont  [1]{#1}%
\providecommand \bibfnamefont [1]{#1}%
\providecommand \citenamefont [1]{#1}%
\providecommand \href@noop [0]{\@secondoftwo}%
\providecommand \href [0]{\begingroup \@sanitize@url \@href}%
\providecommand \@href[1]{\@@startlink{#1}\@@href}%
\providecommand \@@href[1]{\endgroup#1\@@endlink}%
\providecommand \@sanitize@url [0]{\catcode `\\12\catcode `\$12\catcode
  `\&12\catcode `\#12\catcode `\^12\catcode `\_12\catcode `\%12\relax}%
\providecommand \@@startlink[1]{}%
\providecommand \@@endlink[0]{}%
\providecommand \url  [0]{\begingroup\@sanitize@url \@url }%
\providecommand \@url [1]{\endgroup\@href {#1}{\urlprefix }}%
\providecommand \urlprefix  [0]{URL }%
\providecommand \Eprint [0]{\href }%
\providecommand \doibase [0]{http://dx.doi.org/}%
\providecommand \selectlanguage [0]{\@gobble}%
\providecommand \bibinfo  [0]{\@secondoftwo}%
\providecommand \bibfield  [0]{\@secondoftwo}%
\providecommand \translation [1]{[#1]}%
\providecommand \BibitemOpen [0]{}%
\providecommand \bibitemStop [0]{}%
\providecommand \bibitemNoStop [0]{.\EOS\space}%
\providecommand \EOS [0]{\spacefactor3000\relax}%
\providecommand \BibitemShut  [1]{\csname bibitem#1\endcsname}%
\let\auto@bib@innerbib\@empty
%</preamble>
\bibitem [{\citenamefont {Tavoularis}(2005)}]{tavoularis2005measurement}%
  \BibitemOpen
  \bibfield  {author} {\bibinfo {author} {\bibfnamefont {S.}~\bibnamefont
  {Tavoularis}},\ }\href@noop {} {\emph {\bibinfo {title} {{Measurements in
  Fluid Mechanics}}}}\ (\bibinfo  {publisher} {Cambridge University Press},\
  \bibinfo {year} {2005})\BibitemShut {NoStop}%
\bibitem [{\citenamefont {Tavakol}\ and\ \citenamefont
  {Holmes}(2016)}]{Tavakol2016}%
  \BibitemOpen
  \bibfield  {author} {\bibinfo {author} {\bibfnamefont {B.}~\bibnamefont
  {Tavakol}}\ and\ \bibinfo {author} {\bibfnamefont {D.~P.}\ \bibnamefont
  {Holmes}},\ }\href {\doibase http://dx.doi.org/10.1063/1.4944331} {\bibfield
  {journal} {\bibinfo  {journal} {Applied Physics Letters}\ }\textbf {\bibinfo
  {volume} {108}} (\bibinfo {year} {2016}),\
  http://dx.doi.org/10.1063/1.4944331}\BibitemShut {NoStop}%
\bibitem [{\citenamefont {Ducloux}\ \emph {et~al.}(2007)\citenamefont
  {Ducloux}, \citenamefont {Talbi}, \citenamefont {Gimeno}, \citenamefont
  {Viard}, \citenamefont {Pernod}, \citenamefont {Preobrazhensky},\ and\
  \citenamefont {Merlen}}]{Ducloux2007}%
  \BibitemOpen
  \bibfield  {author} {\bibinfo {author} {\bibfnamefont {O.}~\bibnamefont
  {Ducloux}}, \bibinfo {author} {\bibfnamefont {A.}~\bibnamefont {Talbi}},
  \bibinfo {author} {\bibfnamefont {L.}~\bibnamefont {Gimeno}}, \bibinfo
  {author} {\bibfnamefont {R.}~\bibnamefont {Viard}}, \bibinfo {author}
  {\bibfnamefont {P.}~\bibnamefont {Pernod}}, \bibinfo {author} {\bibfnamefont
  {V.}~\bibnamefont {Preobrazhensky}}, \ and\ \bibinfo {author} {\bibfnamefont
  {A.}~\bibnamefont {Merlen}},\ }\href {\doibase
  http://dx.doi.org/10.1063/1.2752530} {\bibfield  {journal} {\bibinfo
  {journal} {Applied Physics Letters}\ }\textbf {\bibinfo {volume} {91}}
  (\bibinfo {year} {2007}),\ http://dx.doi.org/10.1063/1.2752530}\BibitemShut
  {NoStop}%
\bibitem [{\citenamefont {Borisenkov}\ \emph {et~al.}(2015)\citenamefont
  {Borisenkov}, \citenamefont {Kholmyansky}, \citenamefont {Krylov},
  \citenamefont {Liberzon},\ and\ \citenamefont {Tsinober}}]{borisenkov2015}%
  \BibitemOpen
  \bibfield  {author} {\bibinfo {author} {\bibfnamefont {Y.}~\bibnamefont
  {Borisenkov}}, \bibinfo {author} {\bibfnamefont {M.}~\bibnamefont
  {Kholmyansky}}, \bibinfo {author} {\bibfnamefont {S.}~\bibnamefont {Krylov}},
  \bibinfo {author} {\bibfnamefont {A.}~\bibnamefont {Liberzon}}, \ and\
  \bibinfo {author} {\bibfnamefont {A.}~\bibnamefont {Tsinober}},\ }\href
  {\doibase 10.1109/JMEMS.2015.2417213} {\bibfield  {journal} {\bibinfo
  {journal} {Journal of Microelectromechanical Systems}\ }\textbf {\bibinfo
  {volume} {24}},\ \bibinfo {pages} {1503} (\bibinfo {year}
  {2015})}\BibitemShut {NoStop}%
\bibitem [{\citenamefont {Ho}\ and\ \citenamefont {Tai}(1998)}]{ho1998micro}%
  \BibitemOpen
  \bibfield  {author} {\bibinfo {author} {\bibfnamefont {C.-M.}\ \bibnamefont
  {Ho}}\ and\ \bibinfo {author} {\bibfnamefont {Y.-C.}\ \bibnamefont {Tai}},\
  }\href@noop {} {\bibfield  {journal} {\bibinfo  {journal} {Annual Review of
  Fluid Mechanics}\ }\textbf {\bibinfo {volume} {30}},\ \bibinfo {pages} {579}
  (\bibinfo {year} {1998})}\BibitemShut {NoStop}%
\bibitem [{\citenamefont {L{\"o}fdahl}\ and\ \citenamefont {Gad-el
  Hak}(1999)}]{lofdahl1999mems}%
  \BibitemOpen
  \bibfield  {author} {\bibinfo {author} {\bibfnamefont {L.}~\bibnamefont
  {L{\"o}fdahl}}\ and\ \bibinfo {author} {\bibfnamefont {M.}~\bibnamefont
  {Gad-el Hak}},\ }\href@noop {} {\bibfield  {journal} {\bibinfo  {journal}
  {Measurement Science and Technology}\ }\textbf {\bibinfo {volume} {10}},\
  \bibinfo {pages} {665} (\bibinfo {year} {1999})}\BibitemShut {NoStop}%
\bibitem [{\citenamefont {Wang}\ \emph {et~al.}(2009)\citenamefont {Wang},
  \citenamefont {Chen}, \citenamefont {Chang}, \citenamefont {Lin},
  \citenamefont {Lin}, \citenamefont {Fu},\ and\ \citenamefont
  {Lee}}]{wang2009mems}%
  \BibitemOpen
  \bibfield  {author} {\bibinfo {author} {\bibfnamefont {Y.-H.}\ \bibnamefont
  {Wang}}, \bibinfo {author} {\bibfnamefont {C.-P.}\ \bibnamefont {Chen}},
  \bibinfo {author} {\bibfnamefont {C.-M.}\ \bibnamefont {Chang}}, \bibinfo
  {author} {\bibfnamefont {C.-P.}\ \bibnamefont {Lin}}, \bibinfo {author}
  {\bibfnamefont {C.-H.}\ \bibnamefont {Lin}}, \bibinfo {author} {\bibfnamefont
  {L.-M.}\ \bibnamefont {Fu}}, \ and\ \bibinfo {author} {\bibfnamefont {C.-Y.}\
  \bibnamefont {Lee}},\ }\href@noop {} {\bibfield  {journal} {\bibinfo
  {journal} {Microfluidics and nanofluidics}\ }\textbf {\bibinfo {volume}
  {6}},\ \bibinfo {pages} {333} (\bibinfo {year} {2009})}\BibitemShut {NoStop}%
\bibitem [{\citenamefont {Stainback}\ and\ \citenamefont
  {Nagabushana}(1993)}]{stainback1993review}%
  \BibitemOpen
  \bibfield  {author} {\bibinfo {author} {\bibfnamefont {P.}~\bibnamefont
  {Stainback}}\ and\ \bibinfo {author} {\bibfnamefont {K.}~\bibnamefont
  {Nagabushana}},\ }\href@noop {} {\bibfield  {journal} {\bibinfo  {journal}
  {Trans. ASME J. Fluids Eng.}\ }\textbf {\bibinfo {volume} {1}},\ \bibinfo
  {pages} {4} (\bibinfo {year} {1993})}\BibitemShut {NoStop}%
\bibitem [{\citenamefont {Gad-El-Hak}(1989)}]{gad1989advances}%
  \BibitemOpen
  \bibfield  {author} {\bibinfo {author} {\bibfnamefont {M.}~\bibnamefont
  {Gad-El-Hak}},\ }\href@noop {} {\emph {\bibinfo {title} {{Advances in Fluid
  Mechanics Measurements}}}}\ (\bibinfo  {publisher} {Springer},\ \bibinfo
  {year} {1989})\BibitemShut {NoStop}%
\bibitem [{\citenamefont {Schmidt}\ \emph {et~al.}(1988)\citenamefont
  {Schmidt}, \citenamefont {Howe}, \citenamefont {Senturia},\ and\
  \citenamefont {Haritonidis}}]{schmidt1988design}%
  \BibitemOpen
  \bibfield  {author} {\bibinfo {author} {\bibfnamefont {M.~A.}\ \bibnamefont
  {Schmidt}}, \bibinfo {author} {\bibfnamefont {R.~T.}\ \bibnamefont {Howe}},
  \bibinfo {author} {\bibfnamefont {S.~D.}\ \bibnamefont {Senturia}}, \ and\
  \bibinfo {author} {\bibfnamefont {J.~H.}\ \bibnamefont {Haritonidis}},\
  }\href@noop {} {\bibfield  {journal} {\bibinfo  {journal} {IEEE Transactions
  on Electron Devices}\ }\textbf {\bibinfo {volume} {35}},\ \bibinfo {pages}
  {750} (\bibinfo {year} {1988})}\BibitemShut {NoStop}%
\bibitem [{\citenamefont {Jiang}\ \emph {et~al.}(1994)\citenamefont {Jiang},
  \citenamefont {Tai}, \citenamefont {Ho}, \citenamefont {Karan},\ and\
  \citenamefont {Garstenauer}}]{jiang1994theoretical}%
  \BibitemOpen
  \bibfield  {author} {\bibinfo {author} {\bibfnamefont {F.}~\bibnamefont
  {Jiang}}, \bibinfo {author} {\bibfnamefont {Y.-C.}\ \bibnamefont {Tai}},
  \bibinfo {author} {\bibfnamefont {C.-M.}\ \bibnamefont {Ho}}, \bibinfo
  {author} {\bibfnamefont {R.}~\bibnamefont {Karan}}, \ and\ \bibinfo {author}
  {\bibfnamefont {M.}~\bibnamefont {Garstenauer}},\ }in\ \href@noop {} {\emph
  {\bibinfo {booktitle} {IEDM'94 International Electron Devices Meeting}}}\
  (\bibinfo {organization} {IEEE},\ \bibinfo {year} {1994})\ pp.\ \bibinfo
  {pages} {139--142}\BibitemShut {NoStop}%
\bibitem [{\citenamefont {K{\'a}lvesten}(1996)}]{kalvesten1996pressure}%
  \BibitemOpen
  \bibfield  {author} {\bibinfo {author} {\bibfnamefont {E.}~\bibnamefont
  {K{\'a}lvesten}},\ }\emph {\bibinfo {title} {Pressure and wall shear stress
  sensors for turbulence measurements}},\ \href@noop {} {Ph.D. thesis},\
  \bibinfo  {school} {KTH} (\bibinfo {year} {1996})\BibitemShut {NoStop}%
\bibitem [{\citenamefont {Jiang}\ \emph {et~al.}(1996)\citenamefont {Jiang},
  \citenamefont {Tai}, \citenamefont {Gupta}, \citenamefont {Goodman},
  \citenamefont {Tung}, \citenamefont {Huang},\ and\ \citenamefont
  {Ho}}]{jiang1996surface}%
  \BibitemOpen
  \bibfield  {author} {\bibinfo {author} {\bibfnamefont {F.}~\bibnamefont
  {Jiang}}, \bibinfo {author} {\bibfnamefont {Y.-C.}\ \bibnamefont {Tai}},
  \bibinfo {author} {\bibfnamefont {B.}~\bibnamefont {Gupta}}, \bibinfo
  {author} {\bibfnamefont {R.}~\bibnamefont {Goodman}}, \bibinfo {author}
  {\bibfnamefont {S.}~\bibnamefont {Tung}}, \bibinfo {author} {\bibfnamefont
  {J.-B.}\ \bibnamefont {Huang}}, \ and\ \bibinfo {author} {\bibfnamefont
  {C.-M.}\ \bibnamefont {Ho}},\ }in\ \href@noop {} {\emph {\bibinfo {booktitle}
  {Micro Electro Mechanical Systems, 1996, MEMS'96, Proceedings. An
  Investigation of Micro Structures, Sensors, Actuators, Machines and Systems.
  IEEE, The Ninth Annual International Workshop on}}}\ (\bibinfo {organization}
  {IEEE},\ \bibinfo {year} {1996})\ pp.\ \bibinfo {pages}
  {110--115}\BibitemShut {NoStop}%
\bibitem [{\citenamefont {Medina}\ \emph {et~al.}(2014)\citenamefont {Medina},
  \citenamefont {Gilat}, \citenamefont {Ilic},\ and\ \citenamefont
  {Krylov}}]{Medina2014323}%
  \BibitemOpen
  \bibfield  {author} {\bibinfo {author} {\bibfnamefont {L.}~\bibnamefont
  {Medina}}, \bibinfo {author} {\bibfnamefont {R.}~\bibnamefont {Gilat}},
  \bibinfo {author} {\bibfnamefont {B.}~\bibnamefont {Ilic}}, \ and\ \bibinfo
  {author} {\bibfnamefont {S.}~\bibnamefont {Krylov}},\ }\href {\doibase
  10.1016/j.sna.2014.10.016} {\bibfield  {journal} {\bibinfo  {journal}
  {Sensors and Actuators, A: Physical}\ }\textbf {\bibinfo {volume} {220}},\
  \bibinfo {pages} {323} (\bibinfo {year} {2014})}\BibitemShut {NoStop}%
\bibitem [{\citenamefont {Kaajakari}(2009)}]{kaajakari2009small}%
  \BibitemOpen
  \bibfield  {author} {\bibinfo {author} {\bibfnamefont {V.}~\bibnamefont
  {Kaajakari}},\ }\href@noop {} {\emph {\bibinfo {title} {{Practical MEMS}}}}\
  (\bibinfo  {publisher} {Small Gear Publishing},\ \bibinfo {year} {2009})\
  pp.\ \bibinfo {pages} {223--229}\BibitemShut {NoStop}%
\bibitem [{\citenamefont {Robert}\ \emph {et~al.}(2003)\citenamefont {Robert},
  \citenamefont {Saias}, \citenamefont {Billard}, \citenamefont {Boret},
  \citenamefont {Sillon}, \citenamefont {Maeder-Pachurka}, \citenamefont
  {Charvet}, \citenamefont {Bouche}, \citenamefont {Ancey},\ and\ \citenamefont
  {Berruyer}}]{robert2003integrated}%
  \BibitemOpen
  \bibfield  {author} {\bibinfo {author} {\bibfnamefont {P.}~\bibnamefont
  {Robert}}, \bibinfo {author} {\bibfnamefont {D.}~\bibnamefont {Saias}},
  \bibinfo {author} {\bibfnamefont {C.}~\bibnamefont {Billard}}, \bibinfo
  {author} {\bibfnamefont {S.}~\bibnamefont {Boret}}, \bibinfo {author}
  {\bibfnamefont {N.}~\bibnamefont {Sillon}}, \bibinfo {author} {\bibfnamefont
  {C.}~\bibnamefont {Maeder-Pachurka}}, \bibinfo {author} {\bibfnamefont
  {P.}~\bibnamefont {Charvet}}, \bibinfo {author} {\bibfnamefont
  {G.}~\bibnamefont {Bouche}}, \bibinfo {author} {\bibfnamefont
  {P.}~\bibnamefont {Ancey}}, \ and\ \bibinfo {author} {\bibfnamefont
  {P.}~\bibnamefont {Berruyer}},\ }in\ \href@noop {} {\emph {\bibinfo
  {booktitle} {TRANSDUCERS, Solid-State Sensors, Actuators and Microsystems,
  12th International Conference on, 2003}}},\ Vol.~\bibinfo {volume} {2}\
  (\bibinfo {organization} {IEEE},\ \bibinfo {year} {2003})\ pp.\ \bibinfo
  {pages} {1714--1717}\BibitemShut {NoStop}%
\bibitem [{\citenamefont {Sibgatullin}\ and\ \citenamefont
  {Krylov}(2013)}]{sibgatullinexcitation}%
  \BibitemOpen
  \bibfield  {author} {\bibinfo {author} {\bibfnamefont {D.}~\bibnamefont
  {Sibgatullin}, \bibfnamefont {T.and~Schreiber}}\ and\ \bibinfo {author}
  {\bibfnamefont {S.}~\bibnamefont {Krylov}},\ }in\ \href@noop {} {\emph
  {\bibinfo {booktitle} {4th ECCOMAS Thematic Conference on Computational
  Methods in Structural Dynamics and Earthquake Engineering}}},\ \bibinfo
  {editor} {edited by\ \bibinfo {editor} {\bibfnamefont {M.}~\bibnamefont
  {Papadrakakis}}, \bibinfo {editor} {\bibfnamefont {V.}~\bibnamefont
  {Papadopoulos}}, \ and\ \bibinfo {editor} {\bibfnamefont {V.}~\bibnamefont
  {Plevris}}}\ (\bibinfo {year} {2013})\BibitemShut {NoStop}%
\bibitem [{\citenamefont {Maluf}\ and\ \citenamefont
  {Williams}(2004)}]{maluf2004introduction}%
  \BibitemOpen
  \bibfield  {author} {\bibinfo {author} {\bibfnamefont {N.}~\bibnamefont
  {Maluf}}\ and\ \bibinfo {author} {\bibfnamefont {K.}~\bibnamefont
  {Williams}},\ }\href@noop {} {\emph {\bibinfo {title} {Introduction to
  Microelectromechanical Systems Engineering}}}\ (\bibinfo  {publisher} {Artech
  House},\ \bibinfo {year} {2004})\BibitemShut {NoStop}%
\bibitem [{\citenamefont {Joe}\ \emph {et~al.}(2012)\citenamefont {Joe},
  \citenamefont {Linzon}, \citenamefont {Adiga}, \citenamefont {Barton},
  \citenamefont {Kim}, \citenamefont {Ilic}, \citenamefont {Krylov},
  \citenamefont {Parpia},\ and\ \citenamefont {Craighead}}]{Joe2012}%
  \BibitemOpen
  \bibfield  {author} {\bibinfo {author} {\bibfnamefont {D.}~\bibnamefont
  {Joe}}, \bibinfo {author} {\bibfnamefont {Y.}~\bibnamefont {Linzon}},
  \bibinfo {author} {\bibfnamefont {V.}~\bibnamefont {Adiga}}, \bibinfo
  {author} {\bibfnamefont {R.}~\bibnamefont {Barton}}, \bibinfo {author}
  {\bibfnamefont {M.}~\bibnamefont {Kim}}, \bibinfo {author} {\bibfnamefont
  {B.}~\bibnamefont {Ilic}}, \bibinfo {author} {\bibfnamefont {S.}~\bibnamefont
  {Krylov}}, \bibinfo {author} {\bibfnamefont {J.}~\bibnamefont {Parpia}}, \
  and\ \bibinfo {author} {\bibfnamefont {H.}~\bibnamefont {Craighead}},\ }\href
  {\doibase 10.1063/1.4720473} {\bibfield  {journal} {\bibinfo  {journal}
  {Journal of Applied Physics}\ }\textbf {\bibinfo {volume} {111}} (\bibinfo
  {year} {2012}),\ 10.1063/1.4720473}\BibitemShut {NoStop}%
\bibitem [{\citenamefont {Tanaka}(2007)}]{tanaka2007industrial}%
  \BibitemOpen
  \bibfield  {author} {\bibinfo {author} {\bibfnamefont {M.}~\bibnamefont
  {Tanaka}},\ }\href@noop {} {\bibfield  {journal} {\bibinfo  {journal}
  {Microelectronic Engineering}\ }\textbf {\bibinfo {volume} {84}},\ \bibinfo
  {pages} {1341} (\bibinfo {year} {2007})}\BibitemShut {NoStop}%
\bibitem [{\citenamefont {Villaggio}(2005)}]{villaggio2005mathematical}%
  \BibitemOpen
  \bibfield  {author} {\bibinfo {author} {\bibfnamefont {P.}~\bibnamefont
  {Villaggio}},\ }\href@noop {} {\emph {\bibinfo {title} {Mathematical models
  for elastic structures}}}\ (\bibinfo  {publisher} {Cambridge University
  Press},\ \bibinfo {year} {2005})\BibitemShut {NoStop}%
\bibitem [{\citenamefont {Nayfeh}\ and\ \citenamefont
  {Emam}(2008)}]{nayfeh2008exact}%
  \BibitemOpen
  \bibfield  {author} {\bibinfo {author} {\bibfnamefont {A.~H.}\ \bibnamefont
  {Nayfeh}}\ and\ \bibinfo {author} {\bibfnamefont {S.~A.}\ \bibnamefont
  {Emam}},\ }\href@noop {} {\bibfield  {journal} {\bibinfo  {journal}
  {Nonlinear Dynamics}\ }\textbf {\bibinfo {volume} {54}},\ \bibinfo {pages}
  {395} (\bibinfo {year} {2008})}\BibitemShut {NoStop}%
\bibitem [{\citenamefont {Glassbrenner}\ and\ \citenamefont
  {Slack}(1964)}]{glassbrenner1964thermal}%
  \BibitemOpen
  \bibfield  {author} {\bibinfo {author} {\bibfnamefont {C.}~\bibnamefont
  {Glassbrenner}}\ and\ \bibinfo {author} {\bibfnamefont {G.~A.}\ \bibnamefont
  {Slack}},\ }\href@noop {} {\bibfield  {journal} {\bibinfo  {journal}
  {Physical Review}\ }\textbf {\bibinfo {volume} {134}},\ \bibinfo {pages}
  {A1058} (\bibinfo {year} {1964})}\BibitemShut {NoStop}%
\bibitem [{\citenamefont {Pelesko}\ and\ \citenamefont
  {Bernstein}(2002)}]{pelesko2002modeling}%
  \BibitemOpen
  \bibfield  {author} {\bibinfo {author} {\bibfnamefont {J.~A.}\ \bibnamefont
  {Pelesko}}\ and\ \bibinfo {author} {\bibfnamefont {D.~H.}\ \bibnamefont
  {Bernstein}},\ }\href@noop {} {\emph {\bibinfo {title} {{Modeling MEMS and
  NEMS}}}}\ (\bibinfo  {publisher} {CRC press},\ \bibinfo {year} {2002})\ pp.\
  \bibinfo {pages} {70--74}\BibitemShut {NoStop}%
\bibitem [{\citenamefont {Mills}(1999)}]{mills1999basic}%
  \BibitemOpen
  \bibfield  {author} {\bibinfo {author} {\bibfnamefont {A.~F.}\ \bibnamefont
  {Mills}},\ }\href@noop {} {\emph {\bibinfo {title} {Basic Heat and Mass
  Transfer}}}\ (\bibinfo  {publisher} {Prentice Hall},\ \bibinfo {year}
  {1999})\BibitemShut {NoStop}%
\bibitem [{\citenamefont {Holman}(1986)}]{holman1986heat}%
  \BibitemOpen
  \bibfield  {author} {\bibinfo {author} {\bibfnamefont {J.~P.}\ \bibnamefont
  {Holman}},\ }\href@noop {} {\emph {\bibinfo {title} {{Heat Transfer}}}}\
  (\bibinfo  {publisher} {McGraw-Hill},\ \bibinfo {year} {1986})\ pp.\ \bibinfo
  {pages} {344--348}\BibitemShut {NoStop}%
\bibitem [{\citenamefont {Whitaker}(1972)}]{whitaker1972forced}%
  \BibitemOpen
  \bibfield  {author} {\bibinfo {author} {\bibfnamefont {S.}~\bibnamefont
  {Whitaker}},\ }\href@noop {} {\bibfield  {journal} {\bibinfo  {journal}
  {AIChE Journal}\ }\textbf {\bibinfo {volume} {18}},\ \bibinfo {pages} {361}
  (\bibinfo {year} {1972})}\BibitemShut {NoStop}%
\bibitem [{\citenamefont {Okada}\ and\ \citenamefont
  {Tokumaru}(1984)}]{okada1984precise}%
  \BibitemOpen
  \bibfield  {author} {\bibinfo {author} {\bibfnamefont {Y.}~\bibnamefont
  {Okada}}\ and\ \bibinfo {author} {\bibfnamefont {Y.}~\bibnamefont
  {Tokumaru}},\ }\href@noop {} {\bibfield  {journal} {\bibinfo  {journal}
  {Journal of Applied Physics}\ }\textbf {\bibinfo {volume} {56}},\ \bibinfo
  {pages} {314} (\bibinfo {year} {1984})}\BibitemShut {NoStop}%
\bibitem [{\citenamefont {Park}\ \emph {et~al.}(2009)\citenamefont {Park},
  \citenamefont {{Mallon Jr}}, \citenamefont {Rastegar},\ and\ \citenamefont
  {Pruitt}}]{park2009review}%
  \BibitemOpen
  \bibfield  {author} {\bibinfo {author} {\bibfnamefont {W.-T.}\ \bibnamefont
  {Park}}, \bibinfo {author} {\bibfnamefont {J.~R.}\ \bibnamefont {{Mallon
  Jr}}}, \bibinfo {author} {\bibfnamefont {A.}~\bibnamefont {Rastegar}}, \ and\
  \bibinfo {author} {\bibfnamefont {B.~L.}\ \bibnamefont {Pruitt}},\
  }\href@noop {} {\bibfield  {journal} {\bibinfo  {journal} {Proceedings of the
  IEEE}\ }\textbf {\bibinfo {volume} {97}},\ \bibinfo {pages} {513} (\bibinfo
  {year} {2009})}\BibitemShut {NoStop}%
\end{thebibliography}%

\end{document}